\documentclass[preprint,10pt,times,twocolumn,a4paper]{elsarticle}

\usepackage{geometry}
\usepackage{graphicx}
\usepackage{amsmath,amssymb}
\usepackage{booktabs}
\usepackage{siunitx}
\usepackage{multirow}
\usepackage{microtype}
\usepackage[hidelinks]{hyperref}
\usepackage{placeins}
\usepackage{stfloats}

\graphicspath{{}}
\usepackage{xcolor}

\newcommand{\nTotal}{476}
\newcommand{\nCompanies}{23}
\newcommand{\nCemI}{311}
\newcommand{\nNonCemI}{165}
\newcommand{\nCompaniesCemI}{20}
\newcommand{\nCompaniesNonCemI}{16}
\newcommand{\nBogue}{283}
\newcommand{\nWDCemI}{292}
\newcommand{\nFullStrength}{468}
\newcommand{\nFullWD}{443}
\newcommand{\nPSDCemI}{200}

\newcommand{\nClassACemI}{55}
\newcommand{\nClassBCemI}{185}
\newcommand{\nClassCCemI}{50}

\begin{document}
\let\arxivPrintFirstPageNotes\printFirstPageNotes
\let\printFirstPageNotes\relax
\twocolumn[{
\begin{frontmatter}

\title{Inference Limits of Routine Cement Characterization for CEM\,I Performance: Evidence From a Multi-Producer Dataset}

\author[deib]{Marchellino Ghorayeb\corref{cor1}}
\cortext[cor1]{Corresponding author}
\ead{marchellino.ghorayeb@uni-weimar.de}
\address[deib]{Data Science in Civil Engineering, Bauhaus-Universität Weimar}

\author[ibm]{Christiane Rößler}
\address[ibm]{Institute for Building Materials, Bauhaus-Universität Weimar}

\author[ibm]{Horst-Michael Ludwig}
\author[deib]{Leon Herrmann}
\author[deib]{Stefan Kollmannsberger}

\begin{abstract}
This study evaluates the prediction of 28-day compressive strength from routine cement characterization across independent producers. Industrial records from 23 European producers cover nearly three decades of testing in a single laboratory, encompassing 476 samples. The archive includes CEM I, CEM II, CEM III, and other blended cements, with the main analysis focused on CEM I. Chemistry and fineness descriptors were evaluated using linear and nonlinear models. Strength class and water demand were also examined as performance labels from standardized testing. Mean absolute errors in 28-day strength prediction were approximately 3 MPa, with linear models performing comparably to nonlinear models. Producer holdout tests assessed transfer to cements from producers excluded from training and yielded absolute errors of the same order. Equivalent alkali showed a consistent negative association with 28-day strength, with K$_2$O carrying most of this signal. Together, these results highlight the value of industrial data collected
under common laboratory conditions for interpretable 28-day strength
prediction, with potential to inform cement design targeting lower
production costs and CO$_2$ emissions.
\end{abstract}


\end{frontmatter}
}]
\let\printFirstPageNotes\arxivPrintFirstPageNotes
\begingroup
\printFirstPageNotes
\endgroup

\section{Introduction}
\label{sec:intro}

Portland cement production is routinely monitored for quality control, generating a continuous stream of data that includes X-ray fluorescence (XRF) oxide chemistry, specific surface area by Blaine air permeability (hereafter Blaine), calculated Bogue phase composition, and particle-size distribution (PSD). Modern cement plants additionally use X-ray diffraction (XRD) with Rietveld refinement analysis to monitor phase assemblage in the clinker. Although EN\,197-1 specifies performance and composition requirements~\cite{en197}, how well routine characterization supports 28-day strength prediction across independent producers remains uncertain. This is relevant for virtual qualification, cement design feedback, and the interpretation of cross-producer performance studies.

Cement compressive strength depends jointly on composition and fineness~\cite{Alexander1972}. Tsivilis and Parissakis~\cite{StrengthModel1995} expressed this dependency as an empirical predictive relationship. Industrial
surveys also confirm that PSD and Blaine contribute beyond bulk
chemistry alone~\cite{PSDStrength1995}. Abdul et al.~\cite{Abdul2025} document substantial
microstructural and chemical variability across industrial
clinkers. Andrade Neto et al.~\cite{AndradeNeto2025} further show, based on 40 industrial clinkers, that reactivity and strength reflect coupled variation in mineralogy, sulfate-alkali ratio,
crystal size, and fineness across plants. What
remains unclear is how much downstream performance
is recoverable from routine measurements across independent
producers within a common analytical framework. In such settings, strong
performance within a restricted data domain does not necessarily imply robust transfer
across shifted material populations~\cite{LiGeneralizability2023}.

Recent reviews document the use of machine learning to predict concrete
mechanical properties from mixture composition and curing
conditions~\cite{BenChaabene2020,Li2022}. Related studies address mortar
compressive strength from mixture parameters~\cite{Asteris2021}. In cement
production, digital control systems are already in use~\cite{Tong2023},
and machine learning has been applied to predict clinker phase composition
from industrial data~\cite{Fayaz2025}. For cement performance, Ramadhan and
Raharjo~\cite{Ramadhan2025} predicted 28-day strength from routine chemical
and physical measurements using data from one producer, leaving transfer
across independent producers untested.

Here, we use standard cement testing data spanning nearly three decades, collected in a single
laboratory using standardized testing procedures. This common experimental basis helps limit
procedural differences that can complicate datasets compiled from
multiple laboratories, supporting more consistent comparison of cement
descriptor--performance relationships.

We use explainable machine learning models as quantitative probes of the
information that bulk chemistry and fineness contain about 28-day strength,
strength class and water demand in CEM\,I. The analysis also examines whether
Blaine and PSD descriptors provide distinct or redundant information.
Because chemistry and process variables covary within plants, producer-holdout
tests assess whether 28-day strength prediction transfers beyond local
production patterns.

The results show useful 28-day strength prediction across independent
producers, while producer-specific variation and the N/R early-strength
designation remain only partially resolved. Unmeasured kiln and grinding variables, together with sulfate speciation, affect
phase assemblage and early hydration kinetics and may contribute to these gaps.

The model presented could support the optimization of cement composition and
fineness to meet multiple performance requirements with fewer experimental
trials. It could also inform future inverse design strategies aimed at
reducing production costs and CO$_2$ emissions while maintaining the
required cement performance.

\section{Materials and Methods}
\label{sec:methods}

\subsection{Dataset collection}
\label{sec:collection}

The database was compiled from cement samples collected and tested over a 27-year period according to the relevant DIN EN cement standards. All cements were classified according to DIN EN\,197-1. Routine cement characterization included density by DIN EN\,196-6, chemical composition by DIN EN\,196-2, Blaine according to DIN EN\,196-6 and particle-size distribution according to EN\,ISO 13320. 
Cement performance was characterized by setting time, water demand, and Le Chatelier soundness according to DIN EN\,196-3, and by strength development according to DIN EN\,196-1. Flexural strength values at 2, 7 and 28 days correspond to three standard
mortar prisms, while compressive strength values correspond to six
determinations on the resulting prism halves tested under reference curing
conditions. 

Although the dataset spans nearly three decades, the core principles of the underlying cement test methods remained stable over this period. In particular, the Blaine method, Vicat-based setting and water-demand procedures, soundness testing and standard mortar-prism strength testing concepts are retained. However, variability is to be expected on an industrial level and for methods involving the oxide measurements and PSD, which are directly tied to minor revisions in plants and changes in measurement equipment.

The compiled database contains \nTotal{} samples from \nCompanies{}
independent European cement producers.
\nCemI{} are CEM\,I (Portland cement) from \nCompaniesCemI{}
producers. The remaining \nNonCemI{} are CEM\,II, CEM\,III, and composite types. Table~\ref{tab:dataset_summary} summarizes the dataset composition.

A smaller supplementary dataset of eight cements was used exclusively for contextual interpretation of hydration and early-strength behavior. This dataset contains selected cements with characterization beyond the main routine quality control database, including isothermal calorimetry and XRD quantitative phase analysis. Calorimetry provides a direct measure of early hydration kinetics, while XRD resolves crystalline clinker and cement phases, including cubic and orthorhombic forms of C$_3$A. These measurements are not used for primary model training, cross-validation, or feature attribution. They are used only to support the discussion of why the N/R early-strength designation is not fully resolved by the routine characterization available in the main database.

\begin{table}[htbp]
\centering
\caption{Breakdown of cement types, number of samples and number of producers.}
\label{tab:dataset_summary}
\begin{tabular}{@{}lrr@{}}
\toprule
Type  & $n$ & Producers \\
\midrule
CEM\,I & \nCemI{} & \nCompaniesCemI{} \\
CEM\,II/III  & \nNonCemI{} & \nCompaniesNonCemI{} \\
\midrule
Total  & \nTotal{} & \nCompanies{} \\
\bottomrule
\end{tabular}
\end{table}

\subsection{Data extraction and preprocessing}
\label{sec:cleaning}

\subsubsection{Excel workbook extraction}

Data were extracted from structured Excel workbooks with one sheet per producer. The features are summarized in Table~\ref{tab:missing}.

All numerical properties are reported in consistent, domain-appropriate units:
compressive and flexural strength in \si{MPa},
Blaine in \si{cm^2\,g^{-1}},
bulk density in \si{g\,cm^{-3}},
particle diameters in \si{\micro\metre},
setting times in hours,
and Le Chatelier soundness (volume stability) in \si{mm}.
Loss on ignition and phase quantities are reported in wt\%, and water demand
as a unit fraction of cement mass. Dimensionless ratios and indices,
including clinker moduli, are reported without units.

Feature completeness varies across the dataset due to its multi-decade and
multi-producer origin, with coverage of individual feature groups summarized in
Table~\ref{tab:missing}.

\subsubsection{Derived feature computation}

Derived cement descriptors were computed from measured oxide composition. This step included equivalent alkali, Bogue phase fractions, and clinker moduli. The Bogue phase fractions were computed from oxide compositions determined according to EN\,196-2~\cite{en196}. They are retained here as standardized descriptors used in plant quality control, not as direct mineralogical phase measurements. Because the Bogue calculations are based on measured oxide concentrations, uncertainties in the oxide measurements propagate into the estimated phase fractions~\cite{Stutzman2014}. For this reason, oxide and Bogue feature sets are treated as alternative descriptors and are not used simultaneously to avoid multicollinearity.

The Bogue calculation from chemical analysis of CEM\,I requires corrections of the calcium content from limestone, free CaO and calcium associated with the set regulator (calcium sulfate).
These calculated Bogue values are corrected descriptors rather than direct phase measurements. Because the received factory cements include additions after clinker production, measured CaO is first corrected for non-clinker calcium contributions before the phase equations are applied.
\begingroup\small
\begin{align}
  [\text{C}_3\text{S}] &=
    4.071\bigl([\text{CaO}] - [\text{CaO}_\text{free}] - 0.7[\text{SO}_3]\bigr)
    \nonumber\\
    &\quad - 7.602[\text{SiO}_2]
    - 6.719[\text{Al}_2\text{O}_3]
    - 1.430[\text{Fe}_2\text{O}_3], \label{eq:c3s}
\end{align}
\endgroup
The remaining Bogue fractions were computed as
\begin{align}
  [\text{C}_2\text{S}]  &= 2.867[\text{SiO}_2] - 0.754[\text{C}_3\text{S}],\\
  [\text{C}_3\text{A}]  &= 2.650[\text{Al}_2\text{O}_3] - 1.692[\text{Fe}_2\text{O}_3],\\
  [\text{C}_4\text{AF}] &= 3.043[\text{Fe}_2\text{O}_3]. \label{eq:c4af}
\end{align}
Here, C$_3$S, C$_2$S, C$_3$A, and C$_4$AF denote the chemical composition of alite, belite, aluminate when foreign elements are not considered. A recent study showed that results for pure, stoichiometric clinker phases
agree more closely with thermodynamic predictions \citep{Hanein2020}.
The correction terms for free lime and sulfate-bound calcium follow common
industry practice, consistent with the treatment of free lime and sulfate
discussed by Taylor~\cite{Taylor1997}. Free CaO is excluded
from the combinable calcium pool, and $0.7[\text{SO}_3]$ approximates the
CaO incorporated into gypsum as CaSO$_4$ ($56.1/80.1 \approx 0.70$).

Negative Bogue values arising at borderline clinker chemistries were clipped to zero, consistent with the non-negativity of phase fractions and common practice in phase estimation. 

The clinker moduli retained as derived descriptors were computed as~\cite{Taylor1997}
\begin{align}
  \text{LSF} &= \frac{[\text{CaO}]}{2.8[\text{SiO}_2]+1.18[\text{Al}_2\text{O}_3]+0.65[\text{Fe}_2\text{O}_3]},
  \label{eq:lsf}\\
  \text{SM} &= \frac{[\text{SiO}_2]}{[\text{Al}_2\text{O}_3]+[\text{Fe}_2\text{O}_3]}, \quad
  \text{AM} = \frac{[\text{Al}_2\text{O}_3]}{[\text{Fe}_2\text{O}_3]}.
  \label{eq:sm_am}
\end{align}
Here, LSF is the lime saturation factor (describing the tendency towards alite-rich compositions), SM is the silica modulus and AM is the alumina modulus. The cubic-to-orthorhombic distribution of the aluminate phase, discussed in Section~\ref{sec:nr_mech}, is more directly associated with alkali incorporation~\cite{Gobbo2004} and cooling history~\cite{Ichikawa1994}. Equivalent alkali was computed as
\begin{equation}
  [\text{Na}_2\text{O}_\text{eq}]
    = [\text{Na}_2\text{O}] + 0.658[\text{K}_2\text{O}].
  \label{eq:alkali}
\end{equation}

One record in which the 7-day compressive strength fell below the 2-day value was retained, as early-age strength was not used as a primary target and all other measured properties were within the observed ranges of the dataset. Producer identities were anonymized and replaced with arbitrary labels prior to analysis.

\subsubsection{PSD data extraction and integration}
\label{sec:psd_extraction}

PSD parameters were not available in the structured Excel workbooks but
were present on paper-format test reports. These certificates were scanned
and digitized, with a locally run Qwen3.5-9B model~\cite{qwen3.5} used as
an extraction aid for the tabulated PSD fields; extracted values were
manually validated against the source reports before integration with the
analytical dataset. A second, smaller dataset containing PSD parameters covered a partial overlap of cement codes. When a code appeared in both sources, the record with the fewest missing values was
retained to maximize feature completeness when adding them to the main dataset.

PSD is described by the Rosin--Rammler (RR) model~\cite{rrs}
\begin{equation}
  R(x) = 1 - \exp\!\left[-(x/d')^n\right],
  \label{eq:rrs}
\end{equation}
where $d'$ (\si{\micro\metre}) is the location parameter and $n$ the
uniformity (slope) exponent.
The directly reported RR parameters are $d_{\text{m}}$ (mean grain size, weighted
arithmetic mean), $d_{\text{mod}}$ (most frequent grain size, i.e., the mode of
the frequency function), $d_{50}$ (median, 50\% passing), $d'$ (location
parameter), and $n$ (slope parameter).
The percentile descriptors $x_{10}$ and $x_{90}$ are not measured
independently but derived analytically by inverting Eq.~\eqref{eq:rrs}, resulting in
\begin{equation}
  x_p = d'\left(-\ln(1-p)\right)^{1/n}, \quad p \in \{0.10,\,0.90\},
  \label{eq:xp}
\end{equation}
computed only for records with $d' > 0$ and $n > 0$ to ensure physically meaningful distributions.
All seven RR descriptors ($d_{\text{m}}$, $d_{\text{mod}}$, $d_{50}$, $d'$, $n$,
$x_{10}$, $x_{90}$) are computed and are ranked individually in the univariate
correlation screen (Section~\ref{sec:corr}). The predictive and SHAP models
(Sections~\ref{sec:shap_results}--\ref{sec:company}) use a four-parameter PSD
subset ($d_{\text{m}}$, $d_{\text{mod}}$, $x_{10}$, and the slope $n$), chosen to
limit redundancy among the strongly collinear percentile descriptors.
PSD parameters are jointly absent for 36\% of CEM\,I records, as PSD measurement
was not routinely conducted for all samples and, when performed, the parameters were recorded separately from the primary dataset. Missing values are handled natively by
tree models or by restricting analyses to the \nPSDCemI{}-sample complete
subset for other models.

Feature coverage after all preprocessing steps is summarized in
Table~\ref{tab:missing}. 

\begin{table}[htbp]
\centering
\caption{Feature group completeness for CEM\,I ($n = \nCemI{}$). For groups containing multiple variables, missingness is reported as the maximum missing count among variables in that group.}
\label{tab:missing}
\begin{tabular}{@{}lrr@{}}
\toprule
Feature group & Missing ($n$) & (\%) \\
\midrule
Blaine / density              &   5 &  1.6 \\
Water demand                  &  19 &  6.1 \\
Compressive / tensile (28-day)   &  21 &  6.8 \\
Oxides                        &  22 &  7.1 \\
Bogue phases / moduli         &  28 &  9.0 \\
Free CaO                      &  28 &  9.0 \\
Soundness                     &  32 & 10.3 \\
LOI                           &  55 & 17.7 \\
Cl (chloride)                 &  71 & 22.8 \\
PSD (all 7 parameters)        & 112 & 36.0 \\
\bottomrule
\end{tabular}
\end{table}

Notably, of the CEM\,I samples, 21 do not have available 28-day compressive strength measurements. The baseline feature set throughout is a mixture of physical features (Blaine, density) mixed with oxides or Bogue phases and equivalent alkali (CEM\,I only), excluding chloride.
PSD augmented results are reported separately on the CEM\,I subset with
\nPSDCemI{} records. For 28-day compressive strength regression, this subset is further reduced by
target availability, complete case filtering, and outlier screening.

\subsection{Analytical models and attribution methods}
\label{sec:models}

The primary performance-inference analyses were carried out on the CEM\,I
subset, unless stated otherwise. This defines the modeling domain before
feature attribution and prediction are evaluated, while the full cement
database is used only in analyses where comparison across cement types is
explicit. The prediction targets are labels associated with standardized testing.
Classification models predict categorical designations, such as the N/R early-strength designation and the 28-day strength class (32.5, 42.5, 52.5). Regression models predict
continuous test measurements, such as water demand and 28-day compressive strength.

Pairwise associations were quantified using Pearson's
correlation coefficient ($r$) for linear association and Spearman's rank
correlation coefficient ($\rho$) for monotonic association. Reporting both
coefficients allows approximately linear trends to be distinguished from more
general monotonic relationships.

Sample-level performance is estimated by 5-fold cross-validation (CV). The
available CEM\,I records are randomly partitioned into five folds of
approximately equal size. In each iteration, four folds are used to fit the
model and the remaining fold is used as an independent test set. This procedure
is repeated five times, with each fold serving as the test set exactly once, so
that every record contributes to the training set in four iterations and to the
test set in one. The reported metrics are $R^2$, mean absolute error (MAE), and within-tolerance
fraction (W.tol.). The mean and standard deviation (SD) are shown across the five test folds.
Because records from the same producers may appear in both training and test
folds, this convention measures prediction for new samples within the
represented producer population, rather than transfer to a previously unseen
producer.

For linear models, features are standardized within each training
fold and the same scaling is applied to the corresponding test fold, so that no
information from the test fold influences the scaling. The ENET penalty strength
and mixing ratio are selected by an inner 5-fold CV performed only on the
training portion of each outer fold. Tree-model hyperparameters were initially
optimized and then held fixed across the reported cross-validation and
producer-holdout analyses. This arrangement keeps the outer test fold fully held
out during ENET parameter selection. Folds are constructed without grouping by
producer. The complementary case, in which an
entire producer is held out from training, is reported separately in
Section~\ref{sec:lopo}.

\subsubsection{Linear regression}

A linear model expresses 28-day compressive strength $y$ as an additive
combination of the input descriptors, meaning that each descriptor
contributes independently to the prediction as follows

\begin{equation}
\hat{y}(\mathbf{x}) = \mathbf{w}^{\top}\mathbf{x} + b,
\label{eq:linear_model}
\end{equation}
where $\mathbf{x}$ is the vector of standardized cement descriptors (Blaine, oxide or Bogue fractions, PSD parameters), $\mathbf{w}$ is the vector of coefficients, and $b$ is the intercept. The coefficients of ordinary least squares (OLS) are obtained by minimizing the squared residuals on the training fold. ENET modifies this objective by adding a combined $\ell_1$ and $\ell_2$ penalty on $\mathbf{w}$,
\begingroup\small
\begin{equation}
\min_{\mathbf{w},b}\;\frac{1}{2N}\sum_{i=1}^{N}\bigl(y_i - \mathbf{w}^{\top}\mathbf{x}_i - b\bigr)^{2} + \alpha\!\left[\rho\,\|\mathbf{w}\|_1 + \tfrac{1-\rho}{2}\,\|\mathbf{w}\|_2^{2}\right],
\label{eq:enet}
\end{equation}
\endgroup
which shrinks small coefficients toward zero and stabilizes the fit when descriptors are correlated. The overall penalty strength $\alpha$ and the mixing ratio $\rho$ are selected by inner 5-fold CV on each training fold.

In this study, linear models serve as a transparent additive reference. They quantify how much of the 28-day strength variation is captured by a sign-consistent, linear combination of routine CEM\,I descriptors. Their standardized coefficients can be read directly as the linear association between each descriptor and strength across producers after accounting for the other descriptors in the model. Oxide and Bogue features are used in separate linear blocks, never simultaneously, due to the structural collinearity introduced by the Bogue calculation (Eqs.~\eqref{eq:c3s}--\eqref{eq:c4af}).

\subsubsection{Tree-based ensemble methods}

A regression tree partitions the descriptor space by a sequence of binary
splits and assigns a constant value to each terminal partition. Random
forest (RF) regression~\cite{Breiman2001} averages predictions over many
trees fitted independently on bootstrap resamples (a same-size sample drawn with replacement) of the training data,
reducing variance through averaging. Gradient boosting instead combines
$M$ trees sequentially, with each new tree fitted to the residuals of the
current ensemble
\begin{equation}
\hat{y}(\mathbf{x}) = \hat{y}_0 + \sum_{m=1}^{M}\eta\,T_m(\mathbf{x};\boldsymbol{\theta}_m),
\label{eq:gbm}
\end{equation}
where each tree $T_m$ targets the residuals of the previous ensemble,
$\eta$ is a learning rate that scales each contribution, and $\hat{y}_0$
is an initial constant set to the mean of the training target (initial prediction before any trees are added). XGBoost
(XGB)~\cite{xgboost} and LightGBM~\cite{lgbm} are widely used machine
learning libraries that implement gradient-boosted trees with second-order
gradient information and tree-level regularization. Their main controlling
parameters, namely the number of trees, the learning rate, the maximum
tree depth, and the minimum number of samples required to form a split,
were initially optimized and then held fixed across the reported analyses.

Where linear models test purely additive descriptor effects, tree ensembles test whether non-linear interactions among the routine descriptors provide additional explanatory power. Both implementations handle missing values natively by routing them along a learned default branch at each split, which allows all \nCemI{} CEM\,I records to enter the model, including the 36\% with missing PSD parameters (Table~\ref{tab:missing}). Tree predictions are insensitive to the structural collinearity of oxide and Bogue features, since splits are evaluated locally on individual descriptors.

\subsubsection{SHAP feature attribution}

SHapley Additive exPlanations (SHAP)~\cite{shap} decompose a model prediction into additive contributions from each input descriptor. These contributions describe the fitted model output and are not interpreted as causal effects of the descriptors on cement performance. For a sample $\mathbf{x}$, the SHAP value $\phi_j$ of descriptor $j$ is its average marginal contribution across all possible subsets $S$ of the remaining descriptors,
\begingroup\small
\begin{equation}
\phi_j = \!\!\sum_{S\subseteq F\setminus\{j\}}\!\!\frac{|S|!\,(|F|-|S|-1)!}{|F|!}\,\bigl[f(S\cup\{j\}) - f(S)\bigr],
\label{eq:shap}
\end{equation}
\endgroup
where $F$ is the full descriptor set and $f(S)$ is the model output conditioned on the descriptors in $S$. The TreeExplainer algorithm exploits the tree structure of XGBoost and LightGBM to evaluate Eq.~\eqref{eq:shap} exactly in polynomial time, in contrast to model-agnostic SHAP variants that rely on sampling.

The mean absolute SHAP value $\overline{|\phi_j|}$ across the samples used to fit the final model is reported as the model's overall reliance on descriptor $j$. For collinear feature blocks, in particular Bogue phases and PSD parameters, individual SHAP values are redistributed across correlated descriptors and become difficult to interpret in isolation. Attributions are therefore aggregated to the group level and reported throughout Section~\ref{sec:shap_results}.

\subsubsection{Evaluation metrics}

The primary metric for model accuracy is MAE (\si{MPa}), with $R^2$ reported
for comparability with the existing literature. The \textit{within-tolerance
fraction} (W-tol.) is additionally reported as the proportion of test samples
whose absolute prediction error is within 10\% of the measured value. This
threshold is inspired by the numerical criterion specified in EN\,196-1 for the six
individual compressive-strength determinations used to calculate a test result:
a determination differing by more than $\pm10\%$ from their mean is discarded
and the mean is recalculated from the remaining five values~\cite{en196compress}.
W-tol. therefore provides a standard-based relative-error band for evaluating
model predictions.

\section{Results and Discussion}
\label{sec:results}

\subsection{Strength distributions and the case for regression}
\label{sec:compliance}

\begin{figure}[htbp]
  \centering
  \includegraphics[width=\linewidth]{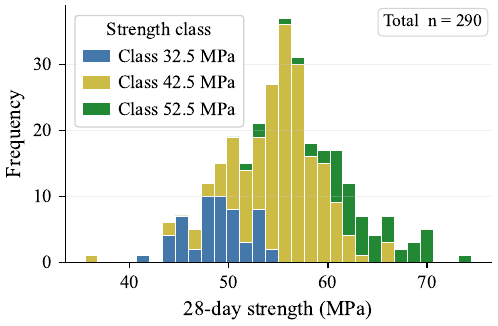}
  \caption{28-day compressive strength distributions overlaid by strength classes for CEM\,I.}
  \label{fig:str_dist}
\end{figure}

Figure~\ref{fig:str_dist} and Table~\ref{tab:sc_stats} reveal three
structural features of the CEM\,I 28-day compressive strength distributions.
First, the standard deviation of 28-day strength is \SIrange{3.1}{4.7}{MPa}
across the three classes, while the strength range within individual classes
extends from approximately \SI{14}{MPa} to over \SI{30}{MPa}, indicating
substantial dispersion within each class.
Second, adjacent classes overlap substantially: the class 32.5 maximum
(\SI{54.3}{MPa}) exceeds the class 42.5 minimum (\SI{35.5}{MPa}) by
\SI{19}{MPa}.
Third, class 32.5 CEM\,I has a mean \SI{16.3}{MPa} above its class minimum,
consistent with producers operating systematically above class minima to maintain a margin for process variability.
Despite this variability, CEM\,I lower bound pass rates for 28-day compressive strength are
100.0\%, 99.5\%, and 98.0\% for classes 32.5, 42.5, and 52.5, respectively.
When upper class limits are also applied, EN\,197-1 class envelope compliance
is 83.6\%, 96.8\%, and 98.0\%. These over-strength margins likely reflect both controlled laboratory testing conditions and producer margins designed to accommodate process and field variability.

\begin{table}[htbp]
\centering
\caption{28-day compressive strength (in MPa) statistics by class, CEM\,I.}
\label{tab:sc_stats}
\begin{tabular}{@{}lrrrrrr@{}}
\toprule
Class & $n$ & Mean & SD & Min & Med. & Max \\
\midrule
32.5 & \nClassACemI{} & 48.8 & 3.1 & 40.7 & 48.9 & 54.3 \\
42.5 & \nClassBCemI{} & 55.5 & 4.1 & 35.5 & 55.8 & 66.1 \\
52.5 & \nClassCCemI{} & 63.2 & 4.7 & 52.3 & 62.8 & 74.5 \\
\bottomrule
\end{tabular}
\end{table}

\begin{figure*}[tbp]
  \centering
  \includegraphics[width=\textwidth]{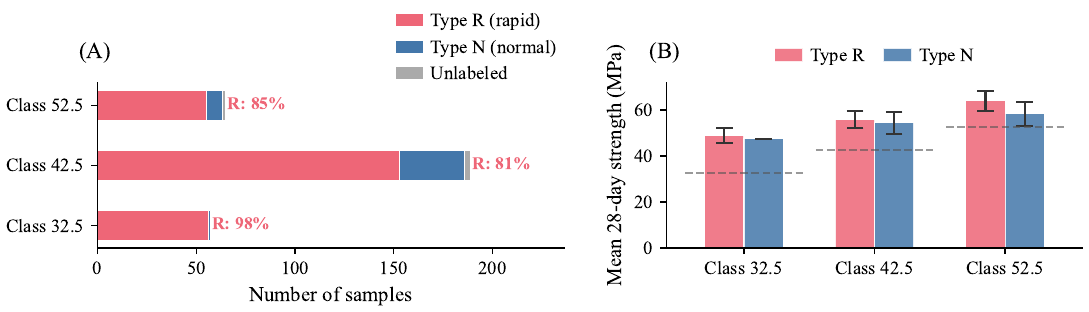}
  \caption{\textbf{(A)} N/R composition per strength class for                CEM\,I.    
           \textbf{(B)} Mean 28-day strength $\pm$1\,SD by designation. Dashed lines indicate EN\,197-1 class minima.}
  \label{fig:nr}
\end{figure*}

Of the \nCemI{} CEM\,I samples, 306 carry an N/R designation; 264
are R-designated and 42 are N-designated. The remaining five samples
are unlabeled and are excluded from N/R analyses.
Class 32.5 is 98\% R-designated, yielding a single N-designated representative and
precluding within-class N/R inference.
Within class 42.5, which is the only class with a meaningful number of N-designated cements
($n=33$), R-designated and N-designated cements exhibit similar mean 28-day compressive strengths
(\SI{55.8}{MPa} and \SI{54.4}{MPa}). This is consistent with the EN\,197-1 distinction between ordinary (N) and high (R) early strength based on early-age requirements
(Fig.~\ref{fig:nr}).

\subsection{Feature statistics and correlation structure}
\label{sec:corr}

Table~\ref{tab:feature_stats} summarizes the primary CEM\,I features. The observed ranges reflect genuine multi-producer variability: Blaine spans a factor of 13 (521--7080\,\si{cm^2\,g^{-1}}), encompassing both
unusually coarse products and finely ground specialty cements; free
CaO ranges from 0 to 6.20\,wt\%, consistent with variation in kiln burning
conditions; and Na$_2$O$_\text{eq}$ spans 0.31--1.85\,wt\% across 20
producers.

\begin{table}[!t]
\centering
\caption{Intrinsic feature summary for CEM\,I ($n=\nCemI{}$). Chemistry in wt\%,
Blaine in \si{cm^2\,g^{-1}}, PSD diameters in \si{\micro\metre}, and
compressive strength in \si{MPa}.}
\label{tab:feature_stats}
\small
\renewcommand{\arraystretch}{0.92}
\setlength{\tabcolsep}{3pt}
\begin{tabular}{@{}lrrrr@{}}
\toprule
Feature & Mean & SD & Min & Max \\
\midrule
\multicolumn{5}{l}{\textit{Physical descriptors}} \\
Blaine & 4063 & 897 & 521 & 7080 \\
$d_{50}$ & 12.4 & 3.9 & 6.0 & 25.9 \\
$d_{\text{mod}}$ & 26.9 & 8.8 & 12.4 & 43.6 \\
\midrule
\multicolumn{5}{l}{\textit{Oxide composition}} \\
CaO & 63.0 & 1.2 & 57.5 & 65.7 \\
SiO$_2$ & 20.4 & 0.8 & 18.2 & 23.6 \\
Al$_2$O$_3$ & 5.0 & 0.7 & 3.2 & 6.8 \\
Fe$_2$O$_3$ & 3.0 & 1.1 & 0.1 & 6.5 \\
SO$_3$ & 3.1 & 0.4 & 1.9 & 3.9 \\
Na$_2$O & 0.20 & 0.10 & 0.00 & 0.90 \\
K$_2$O & 1.00 & 0.30 & 0.30 & 1.80 \\
MgO & 1.40 & 0.60 & 0.10 & 3.60 \\
TiO$_2$ & 0.20 & 0.10 & 0.00 & 0.40 \\
\midrule
\multicolumn{5}{l}{\textit{Derived descriptors}} \\
Na$_2$O$_\text{eq}$ & 0.85 & 0.25 & 0.31 & 1.85 \\
Free CaO & 0.93 & 0.69 & 0.00 & 6.20 \\
C$_3$S & 47.6 & 8.7 & 0.0 & 69.2 \\
C$_2$S & 29.7 & 7.6 & 10.7 & 81.3 \\
C$_3$A & 9.3 & 3.4 & 0.0 & 15.8 \\
C$_4$AF & 10.2 & 3.5 & 0.3 & 22.0 \\
\midrule
Compressive strength (28-day) & 55.6 & 5.9 & 35.5 & 74.5 \\
\bottomrule
\end{tabular}
\end{table}

\begin{figure*}[!t]
  \centering
  \includegraphics[width=0.92\textwidth,height=0.62\textheight,keepaspectratio]{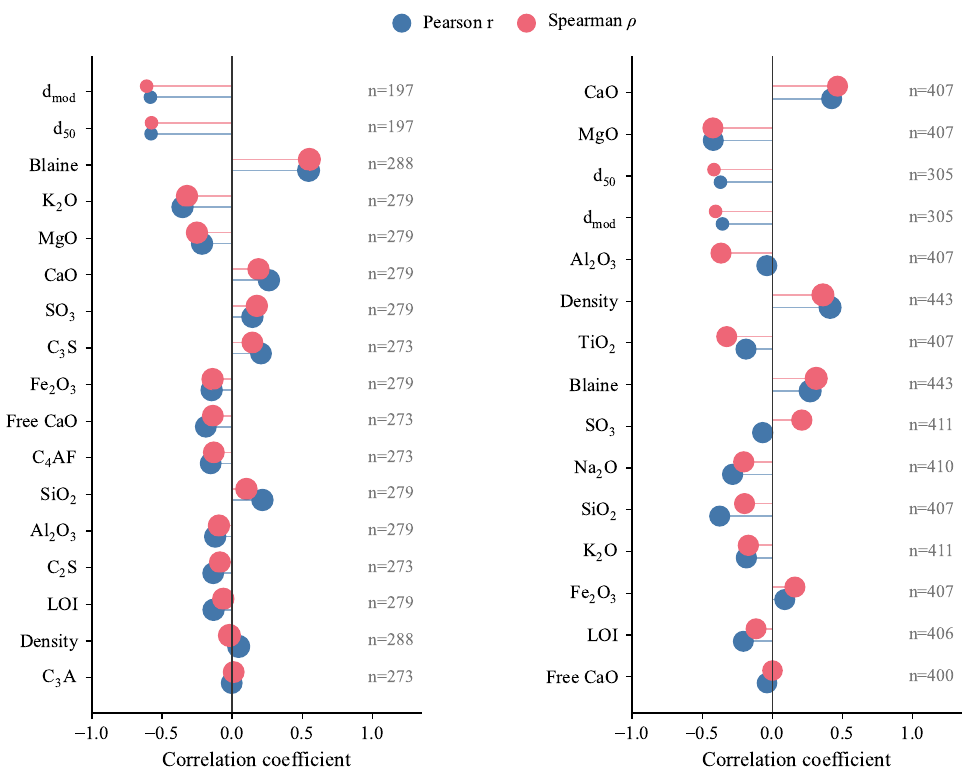}
  \caption{Ranked Spearman $\rho$ and Pearson $r$ correlation coefficients with 28-day compressive
           strength. Left: CEM\,I. Right: CEM\,I, CEM\,II, and CEM\,III. Dot size is proportional to the available sample counts, with $n$ giving the exact sample size;
           line segments connect the two coefficients per feature.}
  \label{fig:corr_ranked}
\end{figure*}

Figure~\ref{fig:corr_ranked} ranks Spearman $\rho$ and Pearson $r$ correlation
coefficients against 28-day strength across all intrinsic features.
Fineness-related descriptors occupy the top of the CEM\,I ranking, in line with
the established dependence of strength and hydration on Blaine
fineness~\cite{Sarkar1990}. The PSD
percentiles $d_{\text{mod}}$ and $d_{50}$ reach $\rho \approx -0.60$ ($n=197$,
pairwise complete with 28-day strength), and Blaine reaches $\rho=0.55$
($n=288$). The sign reversal reflects that finer powders give smaller
percentile diameters and higher specific surface, so the three descriptors
encode the same underlying fineness signal in opposite-signed forms. The near
equality of Pearson and Spearman coefficients indicates near-linear monotonic
relationships over the observed fineness range.
Frigione and Marra~\cite{Frigione1976} and \v{S}kv{\'a}ra et al.~\cite{Skvara1981}
showed that compressive strength can vary with PSD at comparable specific
surface, and Bentz et al.~\cite{Bentz1999} extended this point to broader
performance properties. Under joint multivariate evaluation in this dataset,
Blaine and small PSD percentile subsets are largely interchangeable in
within-distribution performance (\ref{app:fineness}), and the multivariate
dominance of Blaine in Section~\ref{sec:shap_results} reflects its broader
sample coverage rather than a unique informational role.
Several chemical descriptors exhibit moderate correlations ($|\rho| \approx
0.2$--0.35), but no single variable approaches the strength of the fineness
group.
This does not imply that chemistry is uninformative. Rather, the chemical signal
is distributed across correlated oxide variables, and stoichiometric coupling
causes individual effects to partially cancel in univariate projections,
limiting the apparent importance of any single descriptor. When all descriptors
are entered jointly into the same model, chemical descriptors collectively
account for a substantial fraction of the mean absolute SHAP attribution,
showing that the fitted model relies on their combined contribution
(Section~\ref{sec:shap_results}).

Among individual oxides in Fig.~\ref{fig:corr_ranked}, K$_2$O shows the
strongest negative association with 28-day strength ($r=-0.35$,
$\rho=-0.32$, $n=279$), with MgO ranking comparably ($\rho \approx -0.30$).
The K$_2$O association is consistent across the observed range
(0.30--1.80\,wt\%). The derived equivalent alkali descriptor
Na$_2$O$_\text{eq}$ (Eq.~\eqref{eq:alkali}), not shown in
Fig.~\ref{fig:corr_ranked}, reaches $r=-0.34$, $\rho=-0.31$ across the same
$n=279$ subset. Because Na$_2$O$_\text{eq}$ is defined as
Na$_2$O + 0.658\,K$_2$O and Na$_2$O is weakly associated with strength in
this population ($r=-0.07$, $\rho=-0.01$), the Na$_2$O$_\text{eq}$ signal
mainly tracks K$_2$O rather than acting as independent evidence separating
Na$_2$O and K$_2$O chemistry. The same K$_2$O-weighted character carries
through to the model coefficients reported in Section~\ref{sec:linear},
where Na$_2$O$_\text{eq}$ in Bogue-based fits and K$_2$O in oxide-based fits
absorb the same population signal. The MgO association is not interpreted further as an independent mechanistic effect. In this multi-producer dataset, MgO can reflect differences in
raw-material sourcing and plant-specific clinker production conditions. With
the available routine descriptors, the observed MgO--strength association
cannot be separated from these correlated producer-level effects.

In the full dataset, CaO and bulk density displace Blaine as the strongest correlates of 28-day strength because they partly encode cement type and clinker fraction, distinguishing CEM\,I from blended cements containing supplementary cementitious materials or other main constituents. Within the CEM\,I subset, clinker content varies over a narrower range, and fineness descriptors (Blaine and PSD percentiles) provide the strongest single-variable associations with strength.

\subsection{N/R designation as an inference-limit case}
\label{sec:nr_mech}

The EN\,197-1 N/R designation is a kinetic label indicating early strength rather than a 28-day
performance class. As shown in Section~\ref{sec:compliance} (Fig.~\ref{fig:nr}),
N- and R-designated cements within the same strength class can reach similar
28-day strength despite different early-strength requirements. The label is
therefore not expected to define a stable separation for models trained
primarily on 28-day performance.

Routine descriptors nevertheless show population-level tendencies consistent with
known production routes to early strength. R-designated cements are, on average,
finer than N-designated cements and show higher K$_2$O in this dataset. These
trends are compatible with the roles of fineness and sulfate--alkali chemistry in
early hydration~\cite{Enders2024,Jawed1978}. The same designation can be reached through
different combinations of grinding, sulfate--alkali balance, clinker phase
assemblage, and plant practice.

\begin{figure*}[!t]
  \centering
  \includegraphics[width=0.86\textwidth,height=0.48\textheight,keepaspectratio]{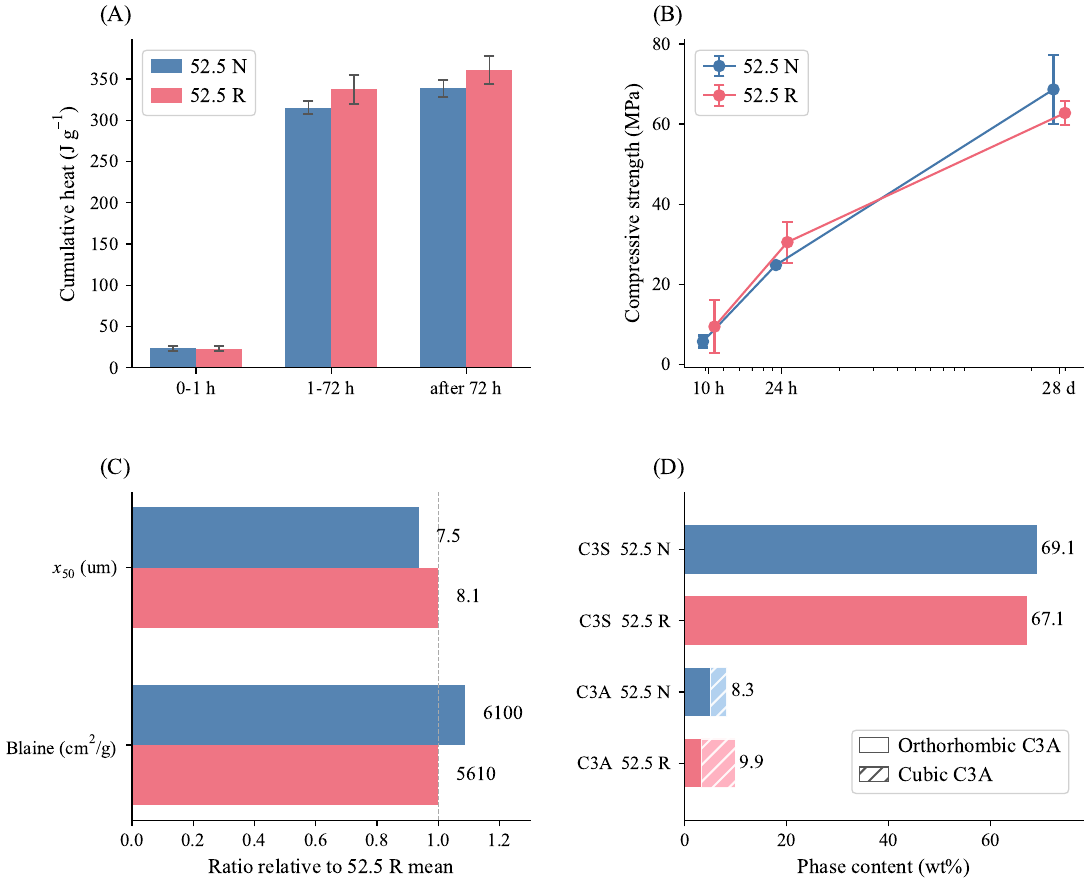}
  \caption{Supplementary dataset, CEM\,I subset ($n=7$): CEM\,I 52.5 N vs.\ CEM\,I 52.5 R.
           (\textbf{A}) Hydration heat values reported for 0--1\,h,
           1--72\,h, and after 72\,h ($w/c=0.75$).
           (\textbf{B}) Compressive strength development at 10h, 24h and 28d with minor $x$-shifts for illustrative purposes.
           (\textbf{C}) Blaine and $d_{50}$, expressed relative to CEM\,I 52.5 R.
           (\textbf{D}) XRD aluminate polymorph distribution (unshaded
           orthorhombic, hatched cubic) within total aluminate content, and total
           alite shown for reference.}
  \label{fig:calor}
\end{figure*}

The supplementary dataset illustrates the importance of mineralogical differences, not captured in routine measurements. In the CEM\,I 52.5 subset ($n=7$; two N and five R), the N-designated cements are finer by
Blaine and $d_{50}$, yet show lower early heat release and lower early strength
than the R-designated cements (Fig.~\ref{fig:calor}); the same comparison also
shows differences in aluminate (C$_3$A) polymorph distribution and alite (C$_3$S)
content. This counterexample points to mineralogical controls not resolved by bulk composition and fineness alone, consistent with the dependence of C$_3$A hydration on polymorphism and solution chemistry~\cite{Myers2017}.

Unsupervised grouping yields similar N/R proportions in both clusters, while supervised classification recovers population level tendencies with substantial overlap (\ref{app:nr_cluster}). N/R therefore illustrates an inference limit: routine descriptors contain useful information about the designation, but do not uniquely resolve kinetic routes influenced by sulfate speciation, aluminate polymorphism and process history.

\subsection{SHAP feature attribution}
\label{sec:shap_results}

Strength class and water demand are performance labels associated with standardized testing and recorded alongside 28-day compressive strength. Hence, before turning to 28-day strength directly, we ask how well these two secondary labels can be inferred from
the intrinsic CEM\,I descriptors, and the degree of information each feature group contributes.

LightGBM~\cite{lgbm} with SHAP TreeExplainer~\cite{shap}
was applied to two secondary targets, strength class
(ordinal) and water demand (continuous) before the primary 28-day strength regression. These analyses assess how much information about the testing labels is recoverable from routine characterization.
Table~\ref{tab:cv_ablation} reports 5-fold CV performance for
the feature subset evaluation.

\begin{figure*}[tbp]
  \centering
  \includegraphics[width=\textwidth]{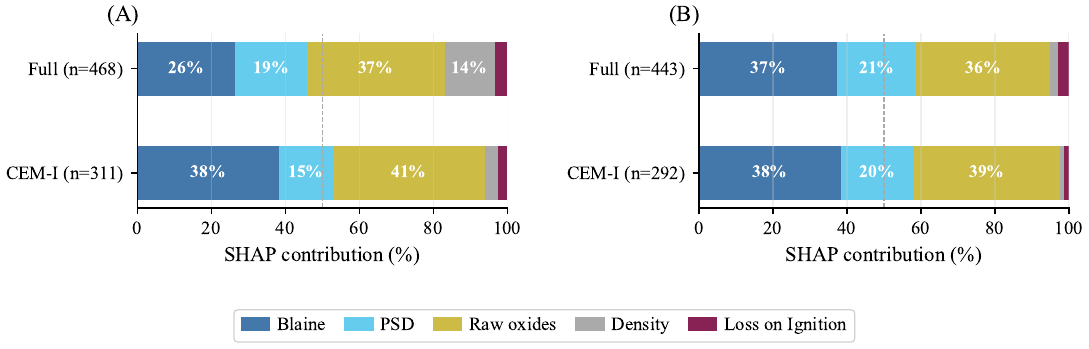}
  \caption{SHAP group contributions for surrogate targets.
           (\textbf{A}) Strength class. (\textbf{B}) Water demand. PSD combines all PSD descriptors.
           Density denotes the bulk density feature.}
  \label{fig:shap_groups}
\end{figure*}

Results in Table~\ref{tab:cv_ablation} show that with all subset features, CEM\,I strength class balanced accuracy (per-class accuracy averaged across classes) reaches
$0.894\pm0.032$, well above the 3-class random reference of $0.33$, and water
demand reaches $R^2=0.713\pm0.049$.
The upper rows and lower block use the same feature definitions, fold
construction, and LightGBM setting. Each lower block model is trained on the
listed feature group or feature group combination.
The percentages are normalized to the all features model within the same
subset experiment and are not additive, because Blaine and PSD contain overlapping information. Blaine alone recovers 84\% of the all features strength class accuracy. Adding
the oxide composition (the chemistry block in these subset experiments) to Blaine
reaches 97\%, after which PSD contributes only modest additional information to
the classification surrogate. This is consistent
with Blaine and small PSD percentile subsets being largely interchangeable
fineness representations in the multivariate models examined here
(\ref{app:fineness}); PSD parameters contribute mainly secondary information
on distribution shape once the dominant fineness signal is already encoded.
This does not imply that Blaine subsumes PSD in general~\cite{Bentz1999}.
Rather, within the present dataset, the choice within the fineness descriptor
group is less consequential than the chemistry-versus-fineness partition
quantified throughout this section.
For water demand, Blaine alone retains 65\% of the $R^2$ obtained with all subset features,
while the oxide composition alone retains 42\%, indicating that fineness remains the
stronger isolated subset. In the SHAP group attribution, however, Blaine and
oxides contribute comparably to water demand. The corresponding
contributions are 38\% and 39\% in CEM\,I, and 37\% and 36\% in the full
cohort, showing that chemistry is not
negligible once the full multivariate model is considered
(Fig.~\ref{fig:shap_groups}).
Within the oxide block, K$_2$O is the leading oxide contribution across
both SHAP studies. This supports the interpretation that, in this population,
the negative equivalent alkali association behaves mainly as a K$_2$O-weighted
signal.

\begin{table}[!t]
\centering
\caption{LightGBM 5-fold CV feature subset performance. Top rows are full feature
references. Parentheses report performance relative to the CEM\,I model using all subset
features and are not additive. Water demand (WD) uses the labeled
water-demand subsets ($\nWDCemI{}$ for CEM\,I and $\nFullWD{}$ for Full).}
\label{tab:cv_ablation}
\small
\renewcommand{\arraystretch}{0.94}
\setlength{\tabcolsep}{3pt}
\resizebox{\columnwidth}{!}{%
\begin{tabular}{@{}lrcc@{}}
\toprule
 & $n$ & Str.\ class (bal.\ acc.) & WD ($R^2$) \\
\midrule
\multicolumn{4}{@{}l}{\textit{All feature reference CV}} \\[1pt]
CEM\,I & \nCemI{} & $0.894\pm0.032$ & $0.713\pm0.049$ \\
Full   & \nFullStrength{} & $0.799\pm0.053$ & $0.699\pm0.045$ \\
\midrule
\multicolumn{4}{@{}l}{\textit{Isolated feature subset performance, CEM\,I}} \\[1pt]
Blaine only         & \nCemI{} & 0.750\ (84\%) & 0.460\ (65\%) \\
Oxides only      & \nCemI{} & 0.681\ (76\%) & 0.297\ (42\%) \\
Blaine\,+\,PSD      & \nCemI{} & 0.794\ (89\%) & 0.484\ (68\%) \\
Blaine\,+\,Oxides    & \nCemI{} & 0.868\ (97\%) & 0.656\ (92\%) \\
All subset features & \nCemI{} & 0.894\ (100\%)  & 0.713\ (100\%)  \\
\bottomrule
\end{tabular}%
}
\end{table}

These results collectively characterize which cement labels are recoverable from routine cement characterization.
Strength class, a coarse but informative interval on 28-day performance, is substantially recoverable (balanced accuracy 0.89 against a random reference of 0.33), and water demand follows closely ($R^2 = 0.71$).
Both strength class and water demand are therefore suitable analytical proxies
for interrogating what routine cement data encodes. The N/R early-strength designation
also shows population level tendencies, with substantial overlap between the two groups (Section~\ref{sec:nr_mech}; \ref{app:nr_cluster}).

\subsection{Predictability limits of 28-day compressive strength}
\label{sec:ml}

\subsubsection{Additive baseline: linear models}
\label{sec:linear}

Linear models constrain the analysis to what bulk cement chemistry, fineness, and Bogue fractions can encode in a direct, additive functional form, providing a transparent additive baseline on performance and a coefficient-level window into feature contributions.
Oxide and Bogue features are used exclusively (not simultaneously) to avoid the structural collinearity of Eqs.~\eqref{eq:c3s}--\eqref{eq:c4af}. The largest oxide variance inflation factor (VIF) for the oxide configuration was Fe$_2$O$_3$ at $8.15$, below the commonly used rule of thumb threshold of 10 for severe multicollinearity, although such thresholds must be interpreted in context~\cite{OBrien2007}. Bogue configurations resulted in varying VIF depending on the configuration, as expected from the stoichiometric dependence of the calculated phases.
For the PSD regression cohort, filtering is sequential. The PSD available CEM\,I cohort contained \nPSDCemI{} rows. Dropping rows with missing features or target values reduced the dataset to 185 rows. A subsequent $5\sigma$ target and standardized feature screen removed four additional rows, each associated with a different feature condition, yielding the final $n=181$ cohort used in Table~\ref{tab:regression}.

Across linear baselines, MAE ranges from 2.73 to 2.86\,\si{MPa}
(Table~\ref{tab:regression}), with 83--91\% of predictions falling within
the $\pm10\%$ relative-error band based on EN\,196-1. This indicates that
routine cement characterization data contain sufficient information to predict
28-day strength with small absolute errors, even in transparent additive models.
Na$_2$O$_\text{eq}$ carries a negative standardized coefficient in all
Bogue-based configurations ($-1.29$ to $-1.34$ across random initializations), while in oxide-based models the corresponding signal is
associated with a strongly negative K$_2$O coefficient (approximately $-2.24$),
with Na$_2$O itself remaining small and slightly positive.
Because Na$_2$O$_\text{eq}$ is defined as Na$_2$O + 0.658\,K$_2$O, and
Na$_2$O is weakly associated with strength in this population, the coefficient
pattern is interpreted as a K$_2$O-weighted alkali signal rather than as
independent evidence separating Na$_2$O and K$_2$O chemistry.
Large residuals persist across all linear configurations, with maximum
absolute residuals of 9--12\,\si{MPa}. Their persistence suggests that part
of the strength variation is not resolved by routine bulk cement descriptors.
Candidate contributing factors, including sulfate--alkali balance, sulfate
form, aluminate polymorph distribution, and clinker reactivity, are discussed
together with the producer-identity decomposition in
Section~\ref{sec:company_onehot}.

\subsubsection{Non-linear feature interactions: tree-based models}
\label{sec:ensemble}

Tree-based models reveal where non-linear chemistry interactions provide
additional recoverable information beyond what linear combinations encode.
For Oxides+PSD, XGB achieves the lowest single MAE across all configurations
(\SI{2.71}{MPa}, $R^2 = 0.638$), narrowly improving on OLS
(\SI{2.80}{MPa}) without changing the qualitative additive picture, while
OLS retains the highest W-tol.\ (91\%). In the Bogue-only configuration,
XGB (\SI{2.97}{MPa}) does not improve on OLS (\SI{2.77}{MPa}). The pattern
across feature sets indicates that the dominant recoverable signal is
additive at this sample size, with non-linear refinements producing modest,
conditional gains rather than a consistent advantage.
A clear pattern emerges when comparing feature sets.
Oxides+PSD achieves the highest $R^2$ across all four models and the overall
best MAE, with XGB at \SI{2.71}{MPa}. Bogue+PSD remains competitive on MAE
under linear models, with OLS at \SI{2.73}{MPa} against \SI{2.80}{MPa} for
Oxides+PSD OLS, while Bogue-only trails on both metrics. This indicates that
the calculated Bogue phase fractions capture much of the strength-relevant
variation present in the oxide composition, but do not preserve all of it.
Adding PSD parameters to Blaine gives only a modest improvement, indicating
that Blaine already captures most of the fineness-related information used by
the models in this dataset, with the detailed PSD descriptors contributing only
limited additional predictive value.

\begin{table}[!tb]
\centering
\caption{5-fold CV performance for 28-day strength, CEM\,I ($n=181$,
         PSD-available subset). W-tol. (fraction within the EN\,196-1
         10\% relative-error band). Inner CV was used for ENET
         hyperparameter selection, while tree-model settings were pre-optimized. Bold values indicate the best value within
         each metric column.}
\label{tab:regression}
\setlength{\tabcolsep}{3.5pt}
\small
\resizebox{\columnwidth}{!}{%
\begin{tabular}{@{}llccc@{}}
\toprule
Feature set & Model & $R^2\pm$SD & MAE$\pm$SD (MPa) & W-tol. \\
\midrule
\multirow{4}{*}{Oxides\,+\,PSD}
  & OLS  & $0.625\pm0.132$ & $2.80\pm0.29$ & \textbf{91\%} \\
  & ENET & $0.615\pm0.108$ & $2.86\pm0.16$ & 90\% \\
  & RF   & $0.579\pm0.098$ & $2.88\pm0.11$ & 86\% \\
  & XGB  & $\mathbf{0.638\pm0.096}$ & $\mathbf{2.71\pm0.20}$ & 89\% \\
\midrule
\multirow{4}{*}{Bogue\,+\,PSD}
  & OLS  & $0.618\pm0.158$ & $2.73\pm0.35$ & 90\% \\
  & ENET & $0.588\pm0.153$ & $2.83\pm0.27$ & 85\% \\
  & RF   & $0.585\pm0.115$ & $2.85\pm0.18$ & 86\% \\
  & XGB  & $0.586\pm0.087$ & $2.90\pm0.22$ & 85\% \\
\midrule
\multirow{4}{*}{Bogue only}
  & OLS  & $0.581\pm0.146$ & $2.77\pm0.27$ & 85\% \\
  & ENET & $0.556\pm0.150$ & $2.85\pm0.29$ & 83\% \\
  & RF   & $0.546\pm0.135$ & $2.94\pm0.27$ & 86\% \\
  & XGB  & $0.546\pm0.104$ & $2.97\pm0.30$  & 84\% \\
\bottomrule
\end{tabular}%
}
\end{table}

\subsection{The chemistry-process boundary: producer identity and measurement limits}
\label{sec:company}
 
The analyses above establish how much information intrinsic chemistry and
fineness contribute in isolation. A complementary question is how much
28-day strength variation remains outside the measured chemistry space, residing in
plant-specific process variables that routine characterization data do not
capture.
Two complementary experiments address this boundary.
Section~\ref{sec:company_onehot} introduces producer identity as one-hot encoded
indicators alongside intrinsic features and decomposes the producer-linked
variance that overlaps with, and lies outside, chemistry and fineness.
Section~\ref{sec:lopo} then removes an entire producer from training and
evaluates transfer, providing a direct test of whether learned
chemistry-to-performance relationships generalize across independent
producers.
 
\subsubsection{Producer identity as an additive indicator}
\label{sec:company_onehot}
 
Producer identity, encoded as one-hot indicator variables (one binary
column per producer, with value 1 for the cement's own producer and 0
elsewhere), alone achieves $R^2 = 0.05$ under both OLS and XGB. This indicates that fixed producer-level offsets explain little of the pooled strength variance when producer identity is used alone. Within the evaluated models, substantially more variance is recovered from chemistry and fineness than from producer-level offsets alone.
 
The conditional experiments summarized in Table~\ref{tab:producer_identity} reveal the structure of the producer signal
more precisely. When producer indicators are appended to Oxides+PSD, OLS improves from
$R^2=0.63$ to $0.70$ and MAE decreases from \SI{2.80}{MPa} to
\SI{2.50}{MPa}; XGB remains essentially unchanged. The increment is therefore
best interpreted as residual additive producer context rather than as a
standalone producer effect.

The diagnostic rows constrain this interpretation. PSD+Company and
Oxides+Company both remain below Oxides+PSD, showing that producer identity
does not replace missing chemistry or fineness descriptors. Instead, the
producer increment likely reflects plant-specific processing choices that are
correlated with, but not explicitly represented by, the routine descriptors.
Grinding practice and grinding aids are one such example because they can alter
grindability, PSD, flow behavior, hydration, and strength without appearing as
oxide variables~\cite{Njiru2023GrindingAids}.

The boundary is therefore conditional: chemistry and fineness carry the
dominant recoverable signal, while producer identity adds a small residual term
only after those descriptors are present.
 
The implication for generalization is limited but clear.
Single-producer studies evaluate models within a constrained chemistry and
process envelope, where local oxide-to-property relationships are
comparatively stable. In materials and concrete machine learning, strong in-domain performance
need not imply robust transfer under shifted data
distributions~\cite{LiGeneralizability2023,Li2022}. Industrial clinker
surveys likewise show that different plants occupy distinct mineralogical
and physicochemical regions of composition space~\cite{Abdul2025}, and
that these differences carry through to clinker reactivity and cement
performance~\cite{AndradeNeto2025}.
For that reason, the present multi-producer evaluation is a stricter test
of transferable signal than a single-plant split. A complementary fully
producer-transfer evaluation is reported in Section~\ref{sec:lopo}.
 
The residual variance unexplained by any combination of chemistry, fineness,
and producer identity is consistent with physicochemical variables that
routine measurements do not directly resolve. Sulfate-alkali balance and sulfate
form distribution are one such class of omitted variables~\cite{Odler1983,Samet1997}.
Gypsum-form variation is another~\cite{Lerch1946,Pourchet2009}. Aluminate (C$_3$A) polymorph ratio
is likewise unresolved by chemical analysis~\cite{Gobbo2004}, and sodium and
sulfate sources further modify how those polymorphs hydrate~\cite{AndradeNeto2022}.
Industrial clinker studies also indicate unresolved contributions from trace
mineralogy and minor-element heterogeneity~\cite{Abdul2025}. Related
effects of alite crystal size, reactivity, and cement
performance are further documented by Andrade Neto
et al.~\cite{AndradeNeto2025}, who in addition showed that
performance cannot be inferred from mineralogy alone~\cite{AndradeNeto2026}.
Cooling history may also contribute through its influence on interstitial
phase character~\cite{Ichikawa1994} and on polymorph distribution~\cite{Gobbo2004}.
These are not incidental omissions. They provide plausible mechanisms by which
industrial clinkers with similar bulk oxide compositions can exhibit
different hydration kinetics and strength development.
The present analysis therefore defines the boundary of inference from routine
characterization rather than the totality of all chemical, crystallographic and physically relevant
descriptors.
 
\begin{table}[htbp]
\centering
\caption{Producer identity experiment, CEM\,I 5-fold CV.
         Feature-set comparisons quantify the extent to which producer
         identity overlaps with, and contributes independently of,
         chemistry and fineness information.}
\label{tab:producer_identity}
\setlength{\tabcolsep}{3.5pt}
\resizebox{\columnwidth}{!}{%
\begin{tabular}{@{}llccc@{}}
\toprule
Feature set & Model & $R^2$ & MAE (\si{MPa}) & W-tol. \\
\midrule
Oxides+PSD         & OLS & 0.625 & 2.80 & 91\% \\
Oxides+PSD         & XGB & 0.638 & 2.71 & 89\% \\
Oxides+PSD+Company & OLS & \textbf{0.702} & \textbf{2.50} & \textbf{94\%} \\
Oxides+PSD+Company & XGB & 0.624 & 2.68 & 88\% \\
\midrule
\midrule
Company only       & OLS & 0.051 & 4.29 & 71\% \\
Company only       & XGB & 0.053 & 4.30 & 70\% \\
\midrule
PSD+Company        & OLS & 0.424 & 3.33 & 83\% \\
PSD+Company        & XGB & 0.338 & 3.52 & 77\% \\
\midrule
Oxides+Company     & OLS & 0.294 & 3.69 & 77\% \\
Oxides+Company     & XGB & 0.280 & 3.59 & 79\% \\

\bottomrule
\end{tabular}%
}
\end{table}
 
\subsubsection{Producer-transfer tests}
\label{sec:lopo}
 
The one-hot experiment quantifies producer-linked variance within a joint
training distribution that already contains samples from every producer.
A stricter complementary test holds out an entire producer from training
and evaluates transfer onto that producer, thereby measuring how well
learned chemistry-to-performance relationships generalize when the target
producer is unseen.
 
For the pooled assessment, every producer with at least five CEM\,I cements in
the common PSD-complete cohort was held out separately. This inclusive threshold
results in 11 producers and 164 held-out samples. For each holdout, the remaining
CEM\,I samples form the training set and the refitted model is evaluated only
on the excluded producer. The held-out predictions are then aggregated to
calculate pooled $R^2$, MAE, and within-tolerance fraction. OLS is held fixed
across feature sets and eligibility threshold. This isolates
the effects of feature representation and producer coverage, while eliminating model
selection noise. Thresholds of $n\geq10$ and $n\geq15$ provide the corresponding
sensitivity analysis (Table~\ref{tab:lopo}). These are nested eligibility sets, meaning that lower threshold groups retain
the producers in the higher-threshold sets. The comparison therefore shows
whether including producers with fewer samples changes the overall transfer
result.

\begin{table}[htbp]
\centering
\caption{Producer holdout performance as a function of the minimum eligible
producer size, CEM\,I. OLS is held fixed across feature sets and thresholds;
pooled metrics are calculated from the concatenated held-out predictions.}
\label{tab:lopo}
\setlength{\tabcolsep}{3pt}
\footnotesize
\resizebox{\columnwidth}{!}{%
\begin{tabular}{@{}ccccccc@{}}
\toprule
Elig. & Prod. & Hold.\ $n$ & Feat. & $R^2$ &
MAE (\si{MPa}) & W-tol. \\
\midrule
\multirow{3}{*}{$n\geq5$} & \multirow{3}{*}{11} & \multirow{3}{*}{164} & Oxides+PSD & 0.483 & 3.20 & 84\% \\
 &  &  & Bogue+PSD & 0.504 & 3.06 & 82\% \\
 &  &  & Bogue & 0.523 & 2.94 & 81\% \\
\midrule
\multirow{3}{*}{$n\geq10$} & \multirow{3}{*}{6} & \multirow{3}{*}{131} & Oxides+PSD & 0.426 & 3.02 & 86\% \\
 &  &  & Bogue+PSD & 0.397 & 2.99 & 85\% \\
 &  &  & Bogue & 0.489 & 2.74 & 85\% \\
\midrule
\multirow{3}{*}{$n\geq15$} & \multirow{3}{*}{2} & \multirow{3}{*}{86} & Oxides+PSD & 0.361 & 2.87 & 90\% \\
 &  &  & Bogue+PSD & 0.416 & 2.69 & 88\% \\
 &  &  & Bogue & 0.537 & 2.40 & 92\% \\
\bottomrule
\end{tabular}%
}
\end{table}

At $n\geq5$, pooled $R^2$ values are $0.48$--$0.52$ and pooled MAE values
are \SIrange{2.94}{3.20}{MPa}. The absolute errors remain close to the
within-distribution values reported in Section~\ref{sec:ensemble}, while the
lower pooled $R^2$ indicates that recovery of strength variation is less stable
when the producer is unseen. Increasing the threshold changes the number of
eligible producers and held-out samples, but does not alter this distinction.
When MAE is recomputed with each producer given equal weight, the difference
from pooled MAE is at most 17.5\% and narrows to 1.9\% at $n\geq15$.

Notably, two producers contain only five held-out samples. At this sample size,
individual $R^2$ is highly sensitive to each observation and is too uncertain
to support producer-specific statistical conclusions. These panels are
therefore retained only in the pooled assessment. Company~7 ($n=58$) and Company~13 ($n=28$) are the only producers
meeting the $n\geq15$ threshold and are therefore the two producer-level
cases discussed in~\ref{app:lopo}. Across all configurations, holdout MAE
falls in $2.35$--$3.14$\,\si{MPa} and within-tolerance fractions are
$79$--$93\%$.

Their individual $R^2$ values vary more than the MAE values because it
compares the model with simply using the mean strength of the held-out
producer for every cement. Strength values within each producer are closely
grouped, so this mean is already a good estimate. Small absolute errors can
therefore still produce a lower $R^2$ when the model does not capture the
smaller differences among that producer's cements.

The detailed results show no consistent benefit from adding the compact PSD
subset or using more complex models; Bogue-only OLS or ENET gives the strongest
result for each producer. The fineness comparison likewise shows that Blaine
and the two-percentile representation transfer similarly
(Table~\ref{tab:fineness_sensitivity}). Overall, absolute errors remain
comparable when producers with fewer samples are included, while results for
individual producers remain more variable.

\FloatBarrier

\section{Conclusions}
\label{sec:conclusions}

Using a dataset spanning nearly three decades, this study demonstrates that routine CEM I characterization supports the prediction of 28-day compressive strength across independent producers and identifies the information limits of these measurements. The main findings are:

(i) Linear models remain competitive with tree ensembles for 28-day strength prediction from chemistry and fineness in this dataset. Blaine and a compact PSD representation provide largely interchangeable fineness information within distribution, and combining them adds little predictive information (\ref{app:fineness}). Strength
class and water demand are also recoverable from routine characterization.

(ii) Among individual chemical descriptors, K$_2$O and the derived equivalent alkali (Na$_2$O$_\text{eq}$) show the most consistent negative association with 28-day strength across correlation and regression analyses. This association should be interpreted as a population-level signal, not as an isolated causal effect.

(iii) Routine descriptors recover population level tendencies in the N/R early-strength designation, with substantial overlap between the two groups. The supplementary counterexamples, in which finer N-designated cements exhibited lower early heat release and 24-hour strength, point to mineralogical controls not resolved by bulk composition and fineness alone.

(iv) The producer-holdout tests yielded pooled mean absolute errors of \SIrange{2.94}{3.20}{MPa} for 28-day strength, comparable to errors within the training distribution. Useful 28-day strength prediction therefore extends to producers excluded from training, although the narrower variation within each unseen producer remains harder to resolve. This residual variation marks the practical inference limit of bulk chemistry and fineness. Resolving it may require sulfate speciation, phase-resolved mineralogy, thermal history, grinding practice and other process descriptors that routine characterization does not fully capture.

\section*{CRediT Author Contribution Statement}

\noindent\textbf{Marchellino Ghorayeb:} Conceptualization, Data curation,
Formal analysis, Investigation, Methodology, Software, Validation,
Visualization, Writing -- original draft, Writing -- review and editing.
\textbf{Christiane Rößler:} Conceptualization, Data curation, Investigation,
Methodology, Writing -- review and editing.
\textbf{Horst-Michael Ludwig:} Writing -- review and editing.
\textbf{Leon Herrmann:} Conceptualization, Data curation, Methodology,
Writing -- review and editing.
\textbf{Stefan Kollmannsberger:} Conceptualization, Data curation, Methodology,
Writing -- review and editing.
 
\section*{Declaration of Competing Interests}
The authors declare that they have no known competing financial interests or personal relationships that could have appeared to influence the work reported in this paper.
 
\section*{Acknowledgements}
The supplementary dataset was the result of a cooperative project funded by the German Federal Ministry for Economic Affairs and Energy (BMWK) under the Central Innovation Programme for SMEs (ZIM), Project No. KK50086081E3. The authors gratefully acknowledge Sonocrete GmbH for providing anonymized data that contributed to this research.  
 
\section*{Data Availability}
The anonymized cement dataset supporting the findings of this study, together with a data dictionary describing the variables and units relevant to processing information, is available in Zenodo. The analysis of the main dataset and scripts are deposited in a separate Zenodo record to provide independent access to the dataset and the scripts used to analyze it. The smaller comparative dataset used is available upon request.

\noindent \textbf{Dataset:} \href{https://doi.org/10.5281/zenodo.21389548} {https://doi.org/10.5281/zenodo.21389548} 

\noindent \textbf{Scripts:} \href{https://doi.org/10.5281/zenodo.21390858} {https://doi.org/10.5281/zenodo.21390858}

\appendix

\setcounter{equation}{0}
\setcounter{figure}{0}
\setcounter{table}{0}

\renewcommand{\theequation}{A.\arabic{equation}}
\renewcommand{\thefigure}{A.\arabic{figure}}
\renewcommand{\thetable}{A.\arabic{table}}
\renewcommand{\theHequation}{A.\arabic{equation}}
\renewcommand{\theHfigure}{A.\arabic{figure}}
\renewcommand{\theHtable}{A.\arabic{table}}

\section{Calcium Oxide Correction}
\label{app:correction}

The standard Bogue calculation treats all CaO reported by chemical analysis as available to
form clinker phases. In an industrial CEM\,I this overstates the combinable
calcium, because part of the measured CaO is held in free lime, in the calcium
sulfate added as a set regulator, and in residual carbonate from limestone
(clinker filler or raw-meal carryover). The corrected combinable calcium is
obtained by removing these three pools,
\begin{equation}
[\text{CaO}]_{\text{eff}} =
  [\text{CaO}] - [\text{CaO}]_{\text{free}}
  - [\text{CaO}]_{\text{sulf}} - [\text{CaO}]_{\text{carb}},
\label{eq:cao_eff}
\end{equation}
where $[\text{CaO}]_{\text{free}}$ is the measured free lime, $[\text{CaO}]_{\text{sulf}}$ is the calcium bound by the set regulator and $[\text{CaO}]_{\text{carb}}$ is from the carbonate content.

The calcium bound by the set regulator is estimated by assuming that all
reported SO$_3$ is present as calcium sulfate, so that
\begingroup\small
\begin{equation}
[\text{CaO}]_{\text{sulf}}
  = \frac{M_{\text{CaO}}}{M_{\text{SO}_3}}\,[\text{SO}_3]
  = \frac{56.08}{80.06}\,[\text{SO}_3]
  \approx 0.700\,[\text{SO}_3].
\label{eq:cao_sulf}
\end{equation}
\endgroup

The carbonate-bound calcium is taken from a measured carbonate content when one
is available from XRD or thermal analysis, as
$[\text{CaO}]_{\text{carb}} = (M_{\text{CaO}}/M_{\text{CaCO}_3})\,[\text{CaCO}_3]
\approx 0.560\,[\text{CaCO}_3]$. For most samples no such measurement exists, and
the carbonate is estimated from the loss on ignition (LOI). A fixed set-regulator
content of 3\,wt\% calcium sulfate dihydrate (CaSO$_4\cdot$2H$_2$O, molar mass
$M_{\text{gyp}}=172.17$) is assumed, its structurally bound water is subtracted
from the LOI, and the remainder is attributed to carbonate decomposition,
\begin{align}
[\text{H}_2\text{O}]_{\text{sulf}}
  &= 3.0\,\frac{2M_{\text{H}_2\text{O}}}{M_{\text{gyp}}}
   = 3.0 \times \frac{36.03}{172.17} \approx 0.628,
   \label{eq:gyp_water}\\
[\text{CO}_2]_{\text{carb}}
  &= \max\!\bigl([\text{LOI}] - [\text{H}_2\text{O}]_{\text{sulf}},\ 0\bigr),
   \label{eq:co2}\\
[\text{CaO}]_{\text{carb}}
  &= \frac{M_{\text{CaO}}}{M_{\text{CO}_2}}\,[\text{CO}_2]_{\text{carb}}
   \approx 1.274\,[\text{CO}_2]_{\text{carb}}.
   \label{eq:cao_carb}
\end{align}
The factor in Eq.~\eqref{eq:cao_carb} is the product of the conversion of evolved
CO$_2$ to CaCO$_3$ ($M_{\text{CaCO}_3}/M_{\text{CO}_2} \approx 2.274$) and of
CaCO$_3$ to CaO ($\approx 0.560$).

Because the calcium correction lowers the oxide total, the Bogue basis is
renormalized before the phase equations are applied. SO$_3$, free lime,
chloride, and the water-soluble alkali fractions are excluded from the basis,
and the remaining oxides ($[\text{CaO}]_{\text{eff}}$, SiO$_2$, Al$_2$O$_3$,
Fe$_2$O$_3$, K$_2$O, Na$_2$O, TiO$_2$, MnO, MgO) are scaled to sum to
100\,wt\%,
\begin{equation}
[X]_{\text{norm}} = 100 \cdot \frac{[X]}{\sum_i [X_i]}.
\label{eq:renorm}
\end{equation}
The renormalized values are substituted into Eqs.~\eqref{eq:c3s}--\eqref{eq:c4af}.
MgO and the alkalis remain in the normalization basis but do not enter the phase
equations, consistent with the classical Bogue method.

\begin{table}[htbp]
\centering
\caption{Per-sample change introduced by the CaO correction in the calculated
Bogue descriptors for CEM\,I ($n=283$ paired cements), as correction minus
legacy estimate (wt\%).}
\label{tab:bogue_correction_effect}
\footnotesize
\begin{tabular}{@{}lrrr@{}}
\toprule
Phase & $\Delta$ mean & $\Delta$ min & $\Delta$ max \\
\midrule
C$_3$S  & $-3.3$ & $-25.9$ & $4.2$ \\
C$_2$S  & $9.7$  & $1.1$   & $34.8$ \\
C$_3$A  & $1.1$  & $0.0$   & $2.2$ \\
C$_4$AF & $1.1$  & $0.0$   & $2.3$ \\
\bottomrule
\end{tabular}
\end{table}

The correction shifts the calculated phases by the per-sample amounts in
Table~\ref{tab:bogue_correction_effect}: alite decreases on average
($-3.3$ wt\%) while belite increases ($+9.7$ wt\%), because part of the
measured CaO is reassigned to non-clinker calcium pools before the Bogue basis is
renormalized.

\begin{table}[htbp]
\centering
\caption{Matched XRD and corrected Bogue phase fractions for
CEM\,I cement samples. Values are XRD/Bogue in wt\%; XRD C$_3$A
combines cubic and orthorhombic aluminate, and XRD C$_2$S is
$\beta$-belite.}
\label{tab:xrd_bogue_compare}
\setlength{\tabcolsep}{3pt}
\footnotesize
\begin{tabular}{@{}lcccc@{}}
\toprule
Sample & C$_3$S & C$_2$S & C$_3$A & C$_4$AF \\
\midrule
Sample~245 & 57.2/55.7 & 10.8/22.3 & 11.7/10.7 & 8.4/8.0 \\
Sample~147 & 65.8/57.3 & 6.1/24.5 & 11.9/6.7 & 6.2/9.1 \\
Sample~110 & 63.5/40.5 & 5.3/37.4 & 12.0/9.7 & 8.2/8.9 \\
Sample~253 & 60.1/47.4 & 3.9/26.1 & 9.9/9.6 & 12.4/12.6 \\
\bottomrule
\end{tabular}
\end{table}

To compare the phase assemblage obtained from the Bogue calculation with experimental data, XRD Rietveld analysis was performed on several CEM\,I cements. Table~\ref{tab:xrd_bogue_compare} highlights the results. For the four complete sample matches, C$_3$A and C$_4$AF are on the same scale as the XRD values. The larger residual difference occurs in the C$_3$S--C$_2$S partition, where XRD gives higher C$_3$S and lower C$_2$S phase fractions, consistent with the idealized phase allocation of the Bogue calculation.

\setcounter{equation}{0}
\setcounter{figure}{0}
\setcounter{table}{0}
\renewcommand{\theequation}{B.\arabic{equation}}
\renewcommand{\thefigure}{B.\arabic{figure}}
\renewcommand{\thetable}{B.\arabic{table}}
\renewcommand{\theHequation}{B.\arabic{equation}}
\renewcommand{\theHfigure}{B.\arabic{figure}}
\renewcommand{\theHtable}{B.\arabic{table}}

\section{Unsupervised and supervised N/R analysis}
\label{app:nr_cluster}

At the population level, R-designated cements are finer than N-designated cements
(mean $d_{\text{mod}}$ \SI{25.6}{} versus \SI{32.9}{\micro\metre} in CEM\,I) and
carry higher K$_2$O (1.02 versus 0.77\,wt\%). Finer grinding and higher alkali are
both routine production levers for early strength, and because they tend to be
used together, neither uniquely marks the N/R designation.

In the supplementary CEM\,I 52.5 subset of Fig.~\ref{fig:calor}, the N group
($n=2$) is finer than the R group ($n=5$) by Blaine (6100 versus
5610\,\si{cm^2\,g^{-1}}) and $d_{50}$ (7.5 versus \SI{8.1}{\micro\metre}), yet
releases less heat (after 72\,h, 338.5 versus \SI{360.9}{J\,g^{-1}}) and develops
less early strength (24\,h, 24.8 versus \SI{30.5}{MPa}). XRD shows the R
group richer in cubic aluminate (6.7 versus 3.3\,wt\%) and the N group richer in
orthorhombic aluminate (5.0 versus 3.2\,wt\%) and alite (69.1 versus
67.1\,wt\%). These observations point to mineralogical controls beyond fineness, although this comparison does not isolate the contributions of individual phases.

\begin{figure}[t]
  \centering
  \includegraphics[width=\linewidth]{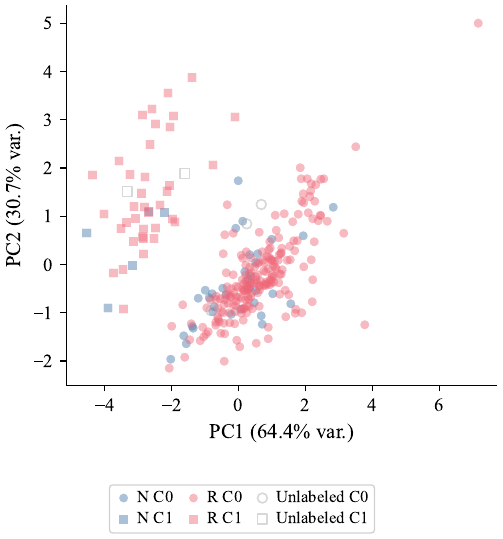}
  \caption{Principal component analysis (PCA) projection of \nBogue{} CEM\,I
           samples using Bogue phase descriptors. Both $k$-means clusters span
           PC1--PC2 (PC1, 64.4\% of the variance, dominated by clinker phase
           content) with similar N/R composition (11.9\% and 14.4\% N). Hollow
           symbols denote samples with no N/R label recorded.}
  \label{fig:nr_cluster}
\end{figure}

The grouping test shows no clear N/R partition in the Bogue phase descriptors. As a first check, the CEM\,I cements were
sorted into two groups using only their
measured composition, with the N/R label hidden from the procedure ($k$-means
clustering). If the measurements distinguished the two early-strength types, one
group would come out mostly N and the other mostly R. They did not: using the
Bogue phases, both groups stayed about 86--88\% R-type (only 11.9\% and 14.4\% of
the cements in the two groups were N-type), essentially the same R-heavy proportion
as the CEM\,I set as a whole (Fig.~\ref{fig:nr_cluster}).

As a complementary check, logistic regression was used to predict each cement's N/R label from routine measurements in the full cement archive. Tenfold stratified cross-validation gave a balanced accuracy of about 75\% (50\% for random guessing). This indicates useful predictive information, with substantial overlap between N- and R-designated cements.

\FloatBarrier

\setcounter{equation}{0}
\setcounter{figure}{0}
\setcounter{table}{0}
\renewcommand{\theequation}{C.\arabic{equation}}
\renewcommand{\thefigure}{C.\arabic{figure}}
\renewcommand{\thetable}{C.\arabic{table}}
\renewcommand{\theHequation}{C.\arabic{equation}}
\renewcommand{\theHfigure}{C.\arabic{figure}}
\renewcommand{\theHtable}{C.\arabic{table}}

\section{Producer holdout robustness}
\label{app:lopo}

Table~\ref{tab:lopo_full} gives the detailed results for the two producers
with at least 15 complete CEM\,I records. It includes all three feature sets
and four models. Training performance is averaged across three random
initializations, while each holdout result uses the same excluded producer.

\begin{table}[htbp]
\centering
\caption{Selected producer holdout results with repeated training
cross-validation, CEM\,I. Training $R^2$ values are mean$\pm$SD across
three random initializations; holdout metrics use the fixed held-out
producer.}
\label{tab:lopo_full}
\setlength{\tabcolsep}{3pt}
\footnotesize
\resizebox{\columnwidth}{!}{%
\begin{tabular}{@{}llcccc@{}}
\toprule
Feat. & Model & Train $R^2\pm$SD & Hold.\ $R^2$ & MAE (\si{MPa}) & W-tol. \\
\midrule
\multicolumn{6}{@{}l}{\textit{Company~7 held out ($n=58$)}} \\[1pt]
\multirow{4}{*}{Oxides+PSD} & OLS & $0.603\pm0.052$ & 0.210 & 2.93 & \textbf{91\%} \\
 & ENET & $0.625\pm0.044$ & 0.404 & 2.55 & \textbf{91\%} \\
 & RF & $0.601\pm0.013$ & 0.386 & 2.55 & 88\% \\
 & XGB & $0.631\pm0.021$ & 0.341 & 2.79 & 88\% \\
\multirow{4}{*}{Bogue+PSD} & OLS & $0.585\pm0.040$ & 0.303 & 2.68 & 88\% \\
 & ENET & $0.578\pm0.023$ & 0.373 & 2.51 & 88\% \\
 & RF & $0.589\pm0.019$ & 0.384 & 2.54 & 90\% \\
 & XGB & $0.605\pm0.016$ & 0.145 & 3.14 & 88\% \\
\multirow{4}{*}{Bogue} & OLS & $0.581\pm0.038$ & 0.456 & 2.42 & \textbf{91\%} \\
 & ENET & $0.555\pm0.035$ & \textbf{0.478} & \textbf{2.35} & \textbf{91\%} \\
 & RF & $0.560\pm0.035$ & 0.292 & 2.61 & 88\% \\
 & XGB & $0.559\pm0.020$ & 0.241 & 2.91 & 88\% \\
\midrule
\multicolumn{6}{@{}l}{\textit{Company~13 held out ($n=28$)}} \\[1pt]
\multirow{4}{*}{Oxides+PSD} & OLS & $0.590\pm0.048$ & 0.560 & 2.76 & 86\% \\
 & ENET & $0.599\pm0.015$ & 0.616 & 2.65 & \textbf{93\%} \\
 & RF & $0.577\pm0.017$ & 0.459 & 3.04 & 82\% \\
 & XGB & $0.585\pm0.003$ & 0.515 & 3.14 & 86\% \\
\multirow{4}{*}{Bogue+PSD} & OLS & $0.596\pm0.034$ & 0.562 & 2.69 & 89\% \\
 & ENET & $0.551\pm0.009$ & 0.545 & 2.80 & 89\% \\
 & RF & $0.581\pm0.003$ & 0.437 & 3.11 & 79\% \\
 & XGB & $0.584\pm0.008$ & 0.438 & 2.99 & 79\% \\
\multirow{4}{*}{Bogue} & OLS & $0.571\pm0.029$ & \textbf{0.640} & \textbf{2.35} & \textbf{93\%} \\
 & ENET & $0.530\pm0.028$ & 0.568 & 2.70 & 89\% \\
 & RF & $0.560\pm0.018$ & 0.482 & 2.91 & 86\% \\
 & XGB & $0.581\pm0.013$ & 0.438 & 3.05 & 86\% \\
\bottomrule
\end{tabular}%
}
\end{table}
\FloatBarrier

The two larger holdouts support the same conclusion as the pooled analysis.
Absolute errors remain similar across feature sets. As explained in
Section~\ref{sec:lopo}, the mean strength of each producer is already a strong
estimate, so small absolute errors can still produce a lower $R^2$.
Bogue-only OLS or ENET gives the strongest result for each producer, while
adding the compact PSD subset does not consistently improve transfer.

Repeating the threshold comparison while allowing the model to change selects
either OLS or ENET in every case and leaves the overall conclusion unchanged.

\setcounter{equation}{0}
\setcounter{figure}{0}
\setcounter{table}{0}
\renewcommand{\theequation}{D.\arabic{equation}}
\renewcommand{\thefigure}{D.\arabic{figure}}
\renewcommand{\thetable}{D.\arabic{table}}
\renewcommand{\theHequation}{D.\arabic{equation}}
\renewcommand{\theHfigure}{D.\arabic{figure}}
\renewcommand{\theHtable}{D.\arabic{table}}

\section{Fineness representation comparison}
\label{app:fineness}

This appendix compares three fineness representations in the same PSD-subset for the CEM\,I regression cohort used in Sections~\ref{sec:ml}
and~\ref{sec:company}: Blaine alone; the two-percentile PSD subset
$d_{\text{mod}}+d_{50}$; and Blaine plus the compact PSD subset used in the
main predictive models ($d_{\text{m}}$, $d_{\text{mod}}$, $x_{10}$, and the
Rosin--Rammler slope $n$). Bulk density is included alongside every
representation. To isolate the fineness representation, the chemistry block, the model (OLS), and five data partitions were held fixed.

\begin{table}[htbp]
\centering
\caption{Sensitivity to fineness representation in the CEM\,I
PSD-complete cohort ($n=181$). CV values use the
same five folds with OLS; holdout values are reported for the two selected
producers.}
\label{tab:fineness_sensitivity}
\setlength{\tabcolsep}{2pt}
\scriptsize
\resizebox{\columnwidth}{!}{%
\begin{tabular}{@{}llccccc@{}}
\toprule
Chemistry & Fineness & CV $R^2$ & CV MAE &
\multicolumn{2}{c}{Holdout $R^2$ / MAE (\si{MPa})} \\
\cmidrule(l){5-6}
 & & & (\si{MPa}) & Company~7 & Company~13 \\
\midrule
Oxides & Blaine & 0.578 & 2.84 & 0.299 / 2.83 & 0.628 / 2.39 \\
Bogue & Blaine & 0.581 & 2.77 & 0.456 / 2.42 & 0.640 / 2.35 \\
Oxides & $d_{\text{mod}}+d_{50}$ & 0.634 & 2.74 & 0.308 / 2.87 & 0.591 / 2.55 \\
Bogue & $d_{\text{mod}}+d_{50}$ & 0.618 & 2.74 & 0.459 / 2.44 & 0.519 / 2.74 \\
Oxides & Blaine+PSD & 0.625 & 2.79 & 0.210 / 2.93 & 0.560 / 2.76 \\
Bogue & Blaine+PSD & 0.618 & 2.73 & 0.303 / 2.68 & 0.562 / 2.69 \\
\bottomrule
\end{tabular}}
\end{table}
\FloatBarrier

Within-distribution $R^2$ values are similar when chemistry is present. In the
selected producer holdouts, Blaine
and $d_{\text{mod}}+d_{50}$ are the strongest representations depending on
the producer, while adding the compact PSD subset to Blaine does not consistently improve
transfer. When producer indicators are added, the within-distribution
interchangeability persists and the producer-identity interpretation of
Section~\ref{sec:company_onehot} is unaffected. The conclusion that the
dominant recoverable fineness signal is represented by simple routine
descriptors is therefore not sensitive to the selected fineness encoding.
 
\bibliography{references}
\end{document}